\documentclass[11pt]{book}
\usepackage{graphicx}                  % This is needed for including figures and graphics
\usepackage{amssymb}
\usepackage{amsmath}
\usepackage{bm}
\numberwithin{equation}{section}
\usepackage{multicol}
\usepackage{array}
\newcolumntype{C}[1]{>{\centering\arraybackslash}m{#1}}
\usepackage{epstopdf}
\usepackage{caption}
\usepackage{subcaption}
\usepackage{mathrsfs}
\usepackage{grffile}
\usepackage{floatrow}
\usepackage[amssymb]{SIunits}
\usepackage{floatflt} 
\usepackage[toc,page]{appendix}
\usepackage[listings,theorems]{tcolorbox}
\usepackage[section]{placeins}
\usepackage{wrapfig}
\usepackage{amsthm}
\usepackage[utf8]{inputenc} 
\usepackage{dsfont}
\usepackage{geometry}
\usepackage{setspace}
\usepackage{physics}
\usepackage{authblk}
\usepackage[round]{natbib}
\usepackage{dsfont}
\usepackage[draft]{hyperref}
\usepackage{titlesec}
\usepackage[font=small,labelfont=bf]{caption}

\titleformat*{\section}{\large\bfseries}
\titleformat*{\subsection}{\normalfont\bfseries}
\titleformat*{\subsubsection}{\normalfont\bfseries}

\newtheorem*{theorem*}{Theorem}

\usepackage{breqn}
\usepackage{thumbpdf}

\DeclareMathAlphabet{\mymathbb}{U}{BOONDOX-ds}{m}{n}

\begin{document}

\setcounter{chapter}{8}
\chapter[(Un)detectability of trajectories]{The (un)detectability of trajectories in pilot-wave theory}
%\addtocontents{toc}{\textit{Johannes Fankhauser}\par}
\bibliographystyle{chicago}
%\maketitle
\title{\textbf{The (un)detectability of trajectories in pilot-wave theory}}
%\emph{Johannes Fankhauser\footnote[1]{{\href{mailto:johannes.j.fankhauser@gmail.com}{\it johannes.j.fankhauser@gmail.com}}}}
%\setcounter{footnote}{0}
\emph{Johannes Fankhauser}

\section{Introduction}

Let's start this chapter with the following question. What does quantum mechanics (QM) state about what is \textit{real}? Unfortunately, about a hundred years after the theory's advent, the answer is still: not much. At best, we have a multitude of proposals in the form of quantum interpretations, some of which try to answer this question while some of which do not. But all come with their problems, and the quantum community is deeply split on which one, if any, to opt for. At worst, up to the present day, the quantum theory continues to be fuelled by confusion about what quantum measurement results supposedly signify.

Despite its tremendous empirical accuracy, one question that the theory does not state is: \textit{What} can be inferred from empirical evidence about the physical system that is measured? In other words, what do quantum measurements signify? And what (if any) information do they convey about the physical system?   

The physical world can be defined using two distinct domains. One domain encompasses empirical data, such as observable elements such as measurement outcomes and records in the environment. This domain constitutes what is directly observable -- the fundamental phenomenological facts. Solely residing within this \textit{manifest} domain does not necessitate any inherent commitment to what exists beyond the tangible aspects of the physical world. The second domain -- the \textit{non-manifest} domain -- encompasses elements, objects, and facts that lie beyond direct observation. These contain entities that exist to which a physical theory relates the manifest elements.\footnote{For example, it indicates when a manifest element, e.g.\ a position measurement record, actually refers to a non-manifest element, e.g.\ the position of a particle.} We may label those elements with the familiar attribute of being `ontological'. The non-manifest domain, furthermore, contains another category, which is termed the nomological sector, i.e.\ the laws of physics determining the dynamics of the ontological sector. 

Despite the above problems of quantum mechanics, some argue that the key lies in giving up altogether the relationship between the manifest and non-manifest domains. Hence, resorting to operationalism or some sort of instrumentalism (see, for instance, Bogen (2017); instrumentalism in \cite{sep-scientific-realism}, and references therein). Others attempt to propose a coherent underpinning by appealing to more ontological commitments (see, for instance, Allori 2023). But, unsurprisingly, a half-hearted attempt or wavering between the two will invariably cause trouble as evidenced in modern literature. 

As it turned out, QM teaches us that what it is that is `observed' is not \textit{a priori} entailed by the use of it. As Einstein aptly noted: what is observable is contingent on the theory (cf.\ Heisenberg 1958).

Appealing to this narrative serves to dispel the proposed experiments discussed in this chapter. As shown later, a naive account of the significance of measurement outcomes in the context of pilot-wave theory serves as a paradigm case for unwarranted claims about the relationship of certain experiments and the physical system. Therefore, insights on what can and cannot be inferred from these experiments have consequences on what is detectable in pilot-wave theory. I examine two proposals: (1) the so-called `weak values' and `weak measurements' with regard to the observation and vindication of Bohmian trajectories. The alleged detectability of particle trajectories is discussed in light of weak \textit{velocity} measurements: The temporal evolution of Bohmian particles is governed by a deterministic dynamics. By construction, however, the individual particle trajectories are underdetermined and generically defy detectability in principle. I discuss two proposals in light of which this lore might possibly be called into question: Firstly, the so-called weak measurements. Due to their characteristic weak coupling between the measurement device and the system under study, they permit the experimental probing of quantum systems without essentially disturbing them. It is therefore natural to think that weak measurements of velocity in particular offer the opportunity to actually observe the particle trajectories. If true, such a claim would not only experimentally demonstrate the incompleteness of QM, but it would also provide support of de Broglie--Bohm theory in its standard form, singling it out from an infinitude of empirically equivalent alternative choices for the particle dynamics. (2) The so-called surrealistic trajectories that were invoked in an attempt to falsify and challenge pilot-wave theory: in this case, an experiment was conceived where the Bohmian predictions for the locations of particle differ from the prediction according to standard quantum theory. Where the particle is measured to reside, i.e.\ its alleged `actual' position -- so the story goes -- conflicts with its position derived from the Bohmian laws of motion. Hence, pilot-wave theory is supposed to provide incorrect results about where particles really travel. 

The discussion offers two insights: I show that weak velocity measurements help to clarify the nature of particle trajectories but not in a robust verificatory sense: they constitute no new arguments, let alone empirical evidence, in favour of standard de Broglie--Bohm theory. One must not naively identify weak and actual positions. This is revealed by a careful reconstruction of the physical arguments on which the description of weak velocity measurements rests. Pilot-wave theory is entirely dispensable for a coherent treatment and interpretation of weak velocity measurements; they receive a natural interpretation within standard QM as observational manifestations of the gradient of the wave function's phase. 

Moreover, a rehearsal of surrealistic trajectories is shown to resemble the same conundrum: what inferences are justified from measurement outcomes to the properties of a particle? The puzzle again resolves by a careful analysis of the significance of measurement results in standard quantum theory and pilot-wave theory.

Thus, the common thread of both instances is the insufficient demarcation of accessible measurement records and what they are supposed to represent ontologically.

This chapter closely follows the author’s previously published research, with some passages reproduced. The discussion on weak velocity measurements in Sections 9.2 and 9.3 is based on \cite{FANKHAUSER-weak-measurements}, while the analysis of surrealistic Bohmian trajectories in Section 9.4 draws on \cite{Fankhauser-thesis}.

\section{Underdetermination in pilot-wave theory}
\label{section:underdetermination}

One of the primary interpretative challenges QM has encountered involves comprehending the nature of entanglement: in the so-called EPR paradox, one widely separates the partners of an entangled pair of particles. They can then no longer interact. Hence, we may, according to Einstein, Podolsky, and Rosen, `without in any way disturbing the system' perform (and expect a well-defined outcome of) a position measurement on one partner and a simultaneous momentum measurement on the other partner (Einstein {\em et al.} 1935, p.\ 777). Prima facie, it looks as if thereby we can bypass the uncertainty relations. However, this raises the question of whether QM in its current form is complete: does every element of physical reality have a counterpart in the description of the QM formalism?  

As is well-known, Einstein was `[...] firmly convinced that the essentially statistical character of contemporary quantum theory is solely to be ascribed to the fact that this [theory] operates with an incomplete description of physical systems' (Einstein 1949, p.\ 666).  

To date, the most elaborate attempt to thus `complete' QM 
dates back to \cite{bohm1952suggested} -- `Bohmian Mechanics' or, in recognition of de Broglie's earlier proposal, `de Broglie--Bohm theory'.\footnote{There exist two \textit{distinct} variants of de Broglie--Bohm theory: the `quantum potential'  school (expounded, e.g.\ by Bohm and Hiley 1993, or Holland 1993), and the `first-order formulation', canonised by D\"urr, Goldstein, Zangh\`i, and their collaborators. The present treatment is only concerned with the latter; `de Broglie--Bohm theory' and 'pilot-wave theory' exclusively refer to this variant.} 

But de Broglie--Bohm theory is not free of problems. From its early days on, a principal objection to it\footnote{For subtleties in the early objections to de Broglie--Bohm theory, we refer to \cite{myrvold2003some}.} targets the unobservability of its particle dynamics. By construction, in de Broglie--Bohm theory, the individual particle trajectories seem to be undetectable \textit{in principle}.\footnote{Note, however, that rejecting equilibrium initial configurations in pilot-wave theory would indeed not only permit experimental tests of the theory, but also lead to a violation of signal-locality (cf., for instance, Valentini (1991, 1992). For caveats on the claim that Bohmian non-equilibrium necessarily leads to signalling, see \cite{Fankhauser-thesis}).} Only their statistical averages are observable. They coincide with the standard quantum mechanical predictions. Thereby, standard de Broglie--Bohm theory achieves empirical equivalence with QM.\footnote{Here, I set aside possible subtleties, see \cite{arageorgis2017bohmian}.}

Empirically, the standard guidance equation thus is not the only option. More precisely, it is not necessary for empirical equivalence with QM. Infinitely many different choices 
\begin{equation}
	\label{altguidanceequation}
	\textbf{\emph{v}}^{\Psi}\mapsto \textbf{\emph{v}}^{\Psi}+ |\Psi|^{-2}\textbf{\emph{j}}
\end{equation} are equally possible for otherwise arbitrary vector fields $\textbf{\emph{j}}$ whose divergence vanishes, $\nabla \cdot \textbf{\emph{j}}=0$. They yield coherent alternative dynamics with distinct particle trajectories, while leaving the predictive-statistical content unaltered (Deotto ad Ghirardi 1998). 

One need not even restrict oneself to deterministic dynamics (an option expressly countenanced by, e.g.\ D\"urr and Teufel 2009, Chapter~1.2): a stochastic dynamics, with $|\Psi|^{-2}\textbf{\emph{j}}$, corresponding to a suitable random variable can also be introduced. As a result, the particles would perform random walks, with the r.h.s. of the integral equation
\begin{equation}
	\nonumber
	\textbf{\emph{Q}}(t)-\textbf{\emph{Q}}(t_0)=\int\limits_{t_0}^{t}\textbf{\emph{v}}^{\Psi}d\tau
\end{equation} containing a diffusion term. A proposal of this type is Nelson Stochastics (see, e.g.\ Goldstein 1987; Bacciagaluppi 2005). In short, de Broglie--Bohm theory's individual particle trajectories are observationally inaccessible by construction.

In consequence, de Broglie--Bohm theory is vastly underdetermined by empirical data: all versions of this theory with guidance equations of the type \eqref{altguidanceequation} are experimentally indistinguishable. Yet, the worlds described by them clearly differ (as shown in Fig. \ref{fig:trac}). 

\newsavebox{\smlmat}% Box to store smallmatrix content
\savebox{\smlmat}{$\textbf{\emph{j}}:=\frac{1}{x^2+y^2}\left(\begin{array}{c} 
		-y\\
		x 
	\end{array}\right)$}

\begin{figure}[ht]
	\centering
	\begin{subfigure}[b]{0.35\textwidth}
		\includegraphics[width=\textwidth]{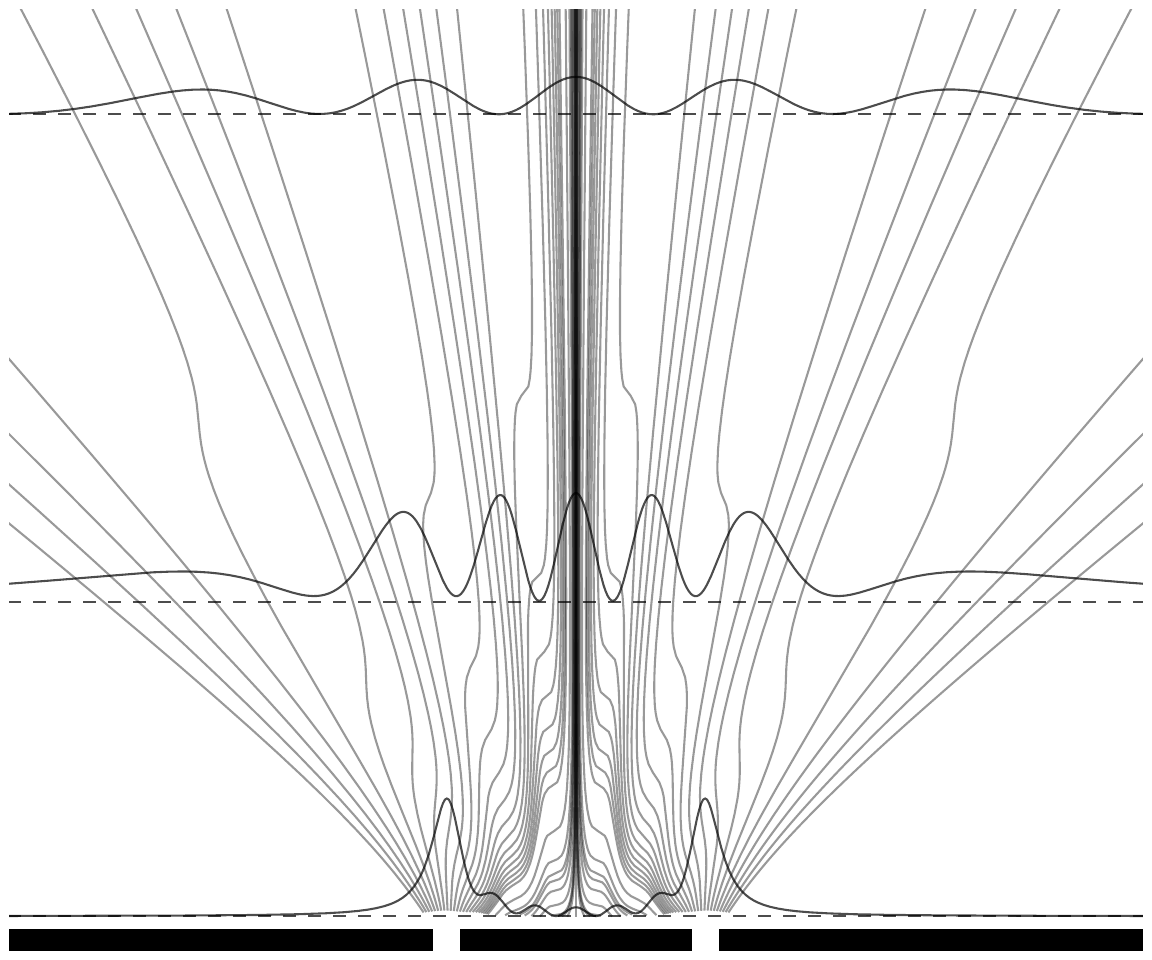}
		\caption{}
		
	\end{subfigure}
	~
	\begin{subfigure}[b]{0.35\textwidth}
		\includegraphics[width=\textwidth]{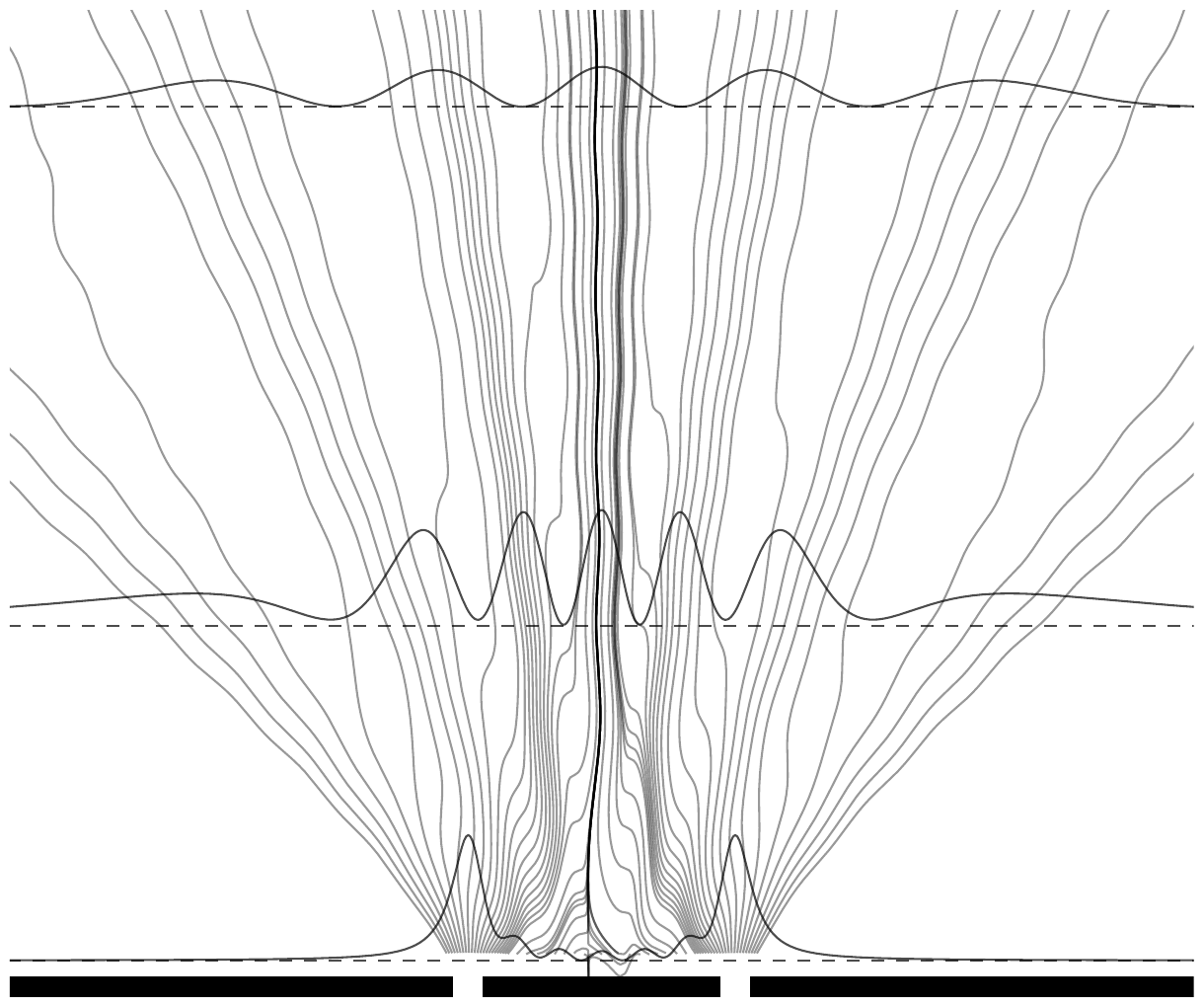}
		\caption{}
		
	\end{subfigure}
	~
	\begin{subfigure}[b]{0.35\textwidth}
		\includegraphics[width=\textwidth]{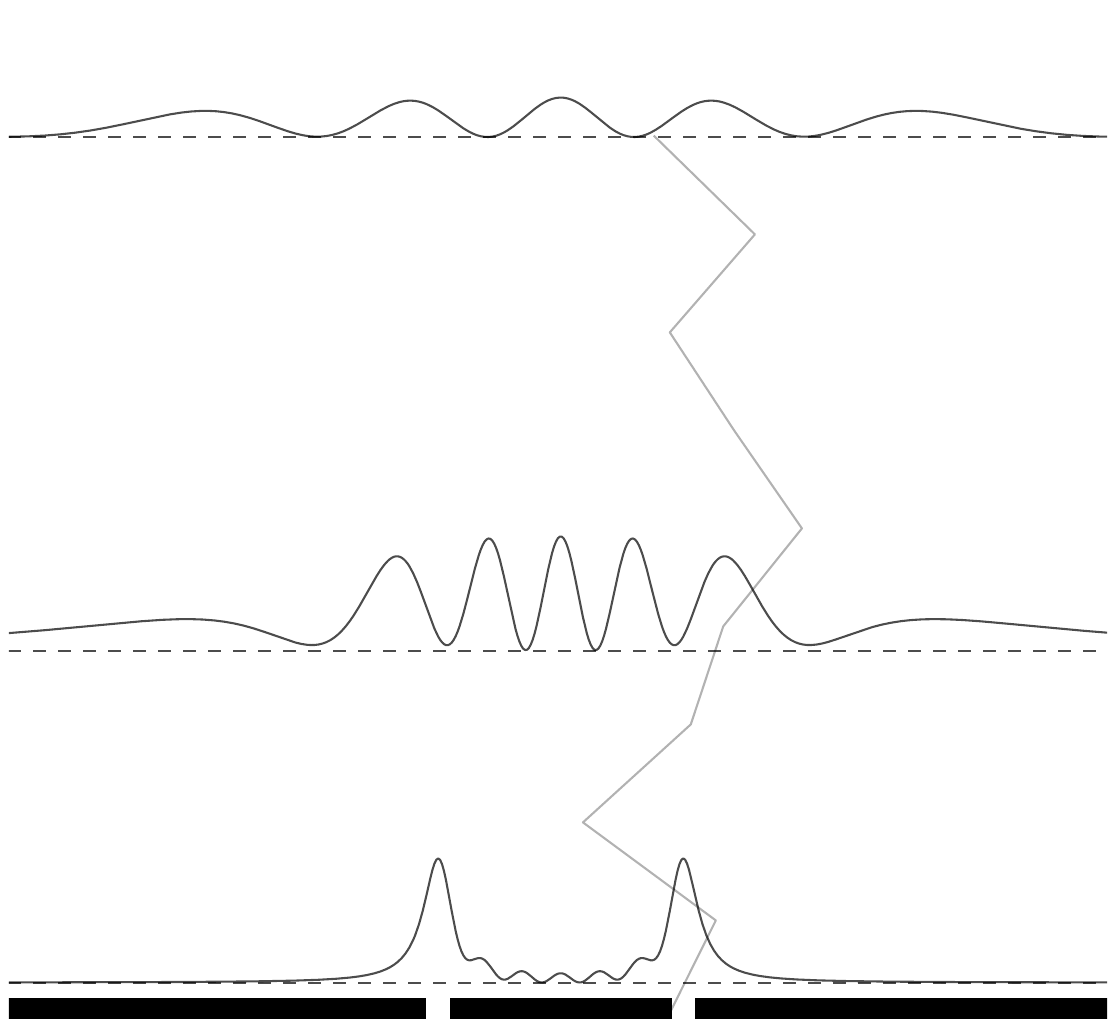}
		\caption{}
		
	\end{subfigure}
	
	\caption{A particle follows different trajectories corresponding to different/non-standard guidance equations.
		(a) The familiar wiggly deterministic trajectories that lead to the interference pattern in a double-slit experiment determined by the standard guidance equation. 
		(b) Alternative trajectories obtained from adding a divergence-free vector field~\usebox{\smlmat} to the standard Bohmian velocity field.  
		(c) A single stochastic trajectory generated by a random variable sampled according to $|\psi|^2$ -- a Bohmian-style theory \textit{without} any particle velocities. For illustration, the probability density $|\psi|^2$ is shown at three different snapshots in time.
		All choices of the dynamics (i.e.\ (a), (b), and (c)) are observationally indiscernible: the resulting measurable distributions at the screen at the top of each figure are the same.}
	\label{fig:trac}	
\end{figure}

This issue renders weak value measurements particularly interesting. By (prima facie) allowing measurements of individual particle trajectories, they appear to directly overcome de Broglie--Bohm theory's underdetermination. But would not that contradict the empirical inaccessibility of the trajectories?

\section{Detecting Bohmian trajectories with weak measurements?}
\label{section:weak measurements}

Methods of weak measurement have opened up a flourishing new field of theoretical and experimental developments  (see, e.g.\ Aharanov and Rohrlich 2005; Tamir
and Cohen 2013; Svensson 2013; Dressel {\em et al}. 2014). Broadly speaking, weak measurements generalize strong measurements in that the final states of measured systems need no longer be eigenstates (and are thus a particular case of positive operator-valued measurements (POVM)). 
Following \cite{aharonov1988result}, weak measurements are measurement processes (modelled via the von Neumann scheme) in which the interaction between the measurement apparatus (`pointer device') and the particle (`system') is weak: it disturbs the wave function only slightly. As a result, one can combine a weak measurement of one quantity (say, initial momenta) and a subsequent ordinary `strong' (or projective) measurement (say, positions).

More precisely, after a weak interaction (say, at $t=0$), the pointer states are not unambiguously correlated with eigenstates of the system under investigation. In contradistinction to strong measurements, the system does not (effectively) `collapse' onto eigenstates; the particles cannot be (say) located very precisely in a single run of an experiment. This apparent shortcoming is compensated for when combined with a strong measurement shortly \textit{after} the weak interaction: the experimenter is then able not only to ascertain the individual particle’s precise location (via the strong measurement), for a sufficiently large ensemble of identically prepared particles with initial state $\psi_{in}$ (viz. Gaussian wave packets with a large spread), she can also gain statistical access to the probability amplitude of all subensembles whose final states -- the so-called `post-selected' state -- have been detected (in the strong measurement) to be $\psi_{fin}$:

\begin{equation}
	\langle \hat{\bf x}\rangle _ w:=\frac{\langle \psi_{fin}|\hat{\bf x}|\psi_{in}\rangle}{\langle\psi_{fin}|\psi_{in}\rangle}.
\end{equation} 

This quantity is called the `weak position value' (for the position operator $\hat{\bf x}$). (The concept is straightforwardly applied to other operators, mutatis mutandis.) It can be shown that after many runs, the pointer’s average position will have shifted by $\langle \hat{\bf x} \rangle_w$.\footnote{In general, $\langle \hat{\bf x} \rangle_w$ is complex, meaning that both the position and the momentum are shifted. Therefore, only the real part of the weak value will be considered in the following discussion.}
Specifically, if we characterize the final/post-selected state via position eigenstates $|\bf x\rangle$, determined in a strong position measurement and unitary evolution of the initial state, we obtain 

\begin{equation}
	\langle \hat{\textbf{x}} (\tau)\rangle _ w=\Re\left(\frac{\langle \textbf{x}|\hat{U}(\tau)\hat{\textbf{x}}|\psi_{in} \rangle}{\langle \textbf{x}|\hat{U}(\tau)|\psi_{in} \rangle}\right),
\end{equation} where $\hat{U}(\tau)$ denotes the unitary time evolution operator during the time interval $[0;\tau]$. Following \cite{wiseman2007grounding}, it is suggestive of construing $\langle \hat{\bf x}(\tau) \rangle _w$ as the mean displacement of particles whose position was found (in a strong position measurement at $t=\tau$) to be at $\bf x$. From this displacement, a natural definition of a velocity field ensues:

\begin{equation}
	\label{operational velocity}
	\textbf{v}(\textbf{x},t)= \lim\limits_{\tau \rightarrow 0}\frac{1}{\tau}(\textbf{x}-\langle\hat{\textbf{x}}(\tau)\rangle_w).
\end{equation} 

Note that all three quantities entering this velocity field --$\tau$, $\bf x$, and $\langle \hat{\textbf{x}} (\tau)\rangle _w$ -- are experimentally accessible. In this sense, the velocity field is `defined operationally' (Wiseman). In what follows, I refer to the application of this measurement scheme --a strong position measurement in short succession upon a particle’s weak interaction with the pointer -- for the associated `operationally defined' velocity field as `Wiseman’s measurement protocol for weak velocity measurements', or simply \textit{`weak velocity measurements'}.

Inspired by \cite{wiseman2007grounding}, eminent Bohmians have advocated such weak measurements as a means of actually observing individual trajectories in standard de Broglie--Bohm theory (e.g.\  Goldstein, 2017, Section 4). Moreover, they point to already performed experiments (e.g.\ Kocsis {\em et al}. 2011; Mahler {\em et al}. 2016) that appear to corroborate de Broglie--Bohm theory's predictions and claim to show the particle trajectories (see also Oianguren-Asua {\em et al}. (2025) for another discussion on weak measurements in the theory). 

For a better understanding of its salient points, let’s now specify such weak velocity measurements in the context of the double-slit experiment. 

\subsection{Weak measurements in the double-slit experiment}
\label{doubleslit} 

Consider the standard double-slit experiment with, say, electrons hitting a screen. It enables the detection of the electrons' positions. This constitutes a strong position measurement. Accordingly, I call this screen the \textit{strong screen}. Let a weak measurement of position be performed between the strong screen and the two slits from which the particles emerge. Let this be called the \textit{weak screen}. The two screens can be moved to measure various distances from the double-slit. Suppose it takes the particles some time $\tau>0$ to travel from the weak screen to the strong screen. 

\begin{figure}[h]
	\centering
	\includegraphics[width=.8\textwidth]{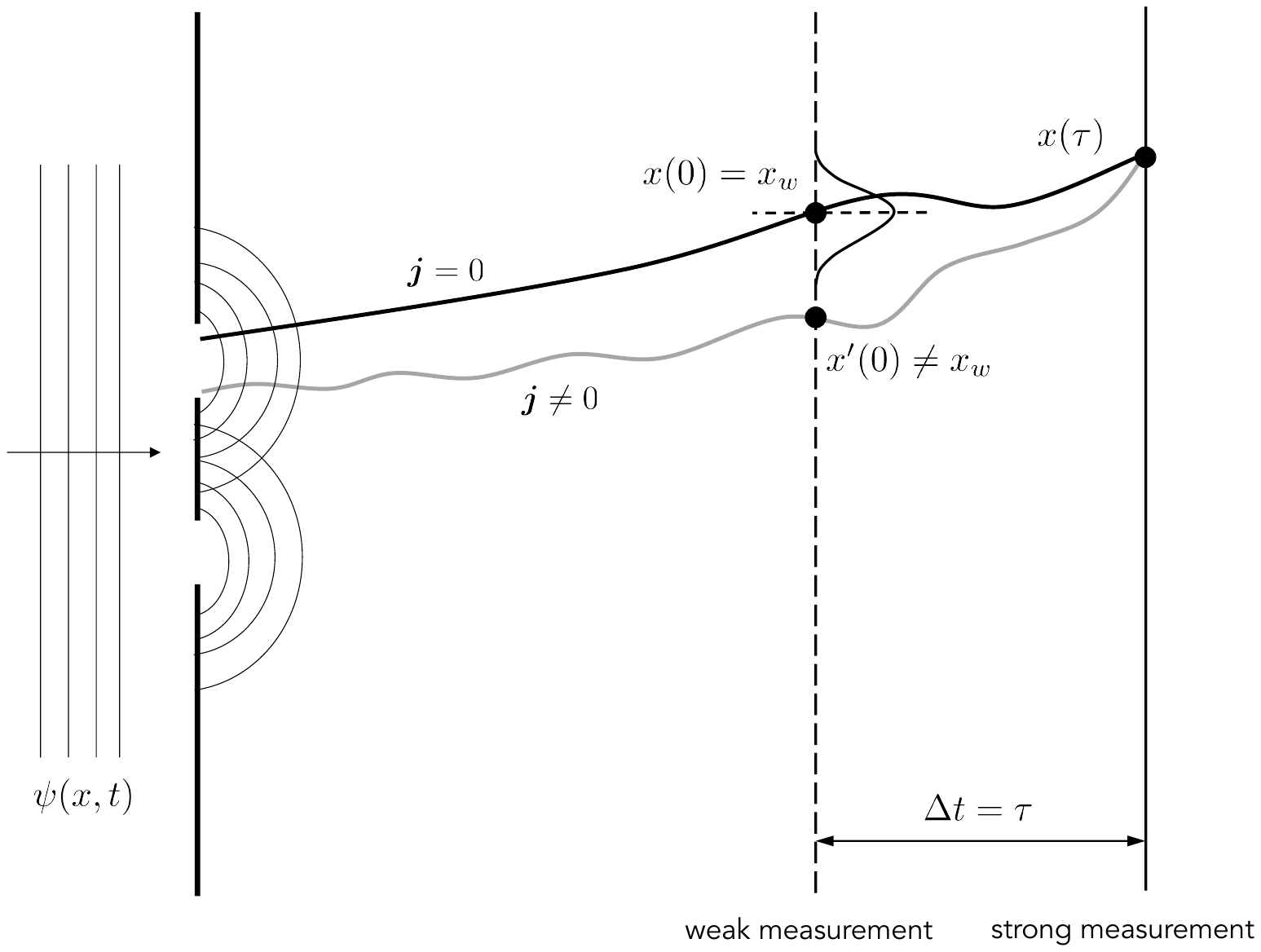}
	\caption{The weak measurement procedure for a given post-selected state $x(\tau)=X_{\tau}$. The weak value is obtained from the distribution on the weak screen. When the velocity field is that of standard de Broglie--Bohm theory ($\textbf{\emph{j}}=0$), the actual position of the particle $x(0)$ matches the weak value $x_w$. For an alternative guidance equation ($\textbf{\emph{j}}\neq 0$), it does not: the particle crosses the weak screen at a point $x'(0)$, other than the weak value. This shows that depending on which guidance equation one chooses, the weak value need not yield the actual position of the particle at time $0$.}
	\label{fig:correspondence}
\end{figure}

After weakly interacting, particle and weak screen are entangled. Hence, only the composite wave function -- \textit{not} the reduced state of the pointer -- evolves unitarily during time $\tau$. Due to this evolution, the post-selected position $x$ on the strong screen will, in general, differ from the weak value $\langle\hat{x}\rangle_w$, obtained from averaging the conditional distribution of the pointer of the weak screen. The procedure is depicted in Fig. \ref{fig:correspondence}). 

On both screens, the wave function is slightly washed out. It differs from an undisturbed state (i.e.\ in the absence of the weak screen). To obtain the two position values -- the weak and the strong one -- strong measurements are now performed both at the weak and the strong screen (i.e.\ on the pointer variable and the target system). For each position outcome $x$ at the strong screen, let's select a sub-ensemble. We then read out the statistical distribution of the position measurement outcomes at the weak screen for any such sub-ensemble. 

We have thus assembled all three observable quantities needed for Wiseman's operationally defined velocity \eqref{operational velocity}: the time $\tau$ that elapsed between the two measurements, the positions $x$ (obtained as values at the strong screen), and the average value of all positions of the sub-ensemble (obtained as values at the weak screen), associated with (i.e.\ post-selected for) $x$.  
This can now be carried out for different positions of the two screens: move them to various locations along the propagation direction, and repeat the same measurement procedure at each position. With this method, one can eventually map the velocity field for a sufficiently large number of measurements.

Kocsis \emph{et al.} have indeed performed an experiment of a similar kind, using weak measurements of momentum. Their result, whose predictions are depicted in Fig. 9.1(a), qualitatively reproduces the trajectories of standard de Broglie--Bohm theory.\footnote{It is worth mentioning that the experiment has been performed for (massless) \textit{photons}. However, standard de Broglie--Bohm theory is a non-relativistic quantum theory for massive particles: it cannot handle photons. The treatment of photons within field-theoretic \textit{extensions} of de Broglie--Bohm theory, capable of dealing with photons (or bosons, more generally), is a delicate matter, which is outside the scope of this chapter. I refer the interested reader to Holland (1993, Chapter~11) and  D\"urr {\em et al}. (2013, Chapter~10) (also for further references). Kocsis \emph{et al.}'s experiment hence has no direct bearing on de Broglie--Bohm theory's status. Rather than the trajectories of \textit{individual} photons, Flack and Hiley (2014, 2016) have argued that Kocsis \emph{et al.}'s experiments measure mean momentum flow lines.} Moreover, it can be shown (cf. the Appendix in Fankhauser and D\"urr 2021) that weak velocity measurements are measurements of the gradient of the phase of the wave function. Thus, they coincide with the definition of standard Bohmian velocities in the guidance equation

\begin{equation}
	\textbf{v}= \frac{\hbar}{m}\nabla S, 
\end{equation} where $S$ is the gradient of the phase of the wave function, $\psi(\textbf{x})=|\psi(\textbf{x})|e^{i S(\textbf{x})}$. 

Note that only the standard quantum mechanical formalism has been utilized for this. Therefore, we may conclude that -- based solely on standard QM -- weak velocity measurements permit experimental access to the gradient of the wave function's phase. 

%\begin{figure}[ht!]
%	\centering
%	\includegraphics[width=0.9\textwidth]{Kocsis trac.jpg}
%	\caption{A weak velocity measurement for photons allows the reconstructions of trajectories, qualitatively identical to those of particles in standard de Broglie--Bohm theory. Particle trajectories in a double-slit experiment performed by \cite{kocsis2011observing}.}
%	\label{fig:Kocsis}
%\end{figure}

\subsection{Why weak velocity measurements do not measure velocities}
\sectionmark{Weak and actual velocities}
\label{weak measurements are not genuine}

Suggestive as these results are, I now show that such measurements could not provide direct experimental evidence displaying the shape of particle trajectories, even if it is assumed that some deterministic particle trajectories exist. They cannot, that is, go anyway to experimentally resolving the underdetermination in putative de Broglie--Bohm theory guidance equations mentioned earlier. I show that a strong assumption is required that would render it question-begging to employ weak velocity measurements in order to infer the particles' \textit{actual} velocities. I argue that this turns out to \textit{presuppose} standard de Broglie--Bohm theory.
\vspace{2mm}

\noindent {\em When do weak and actual velocities coincide?}
%\label{section when do weak and actual velocities coincide}
In the following, $x$ and $y$ denote the position variables of the individual particles to be measured and the measurement apparatus, respectively. For simplicity, consider one dimension only. Let the particles be prepared in the initial state 
\begin{equation}
	|\psi\rangle=\int dx ~\psi(x)|x\rangle.
\end{equation}  

Furthermore, let the pointer device (i.e.\ the weak screen of the double-slit version of weak measurements in Section \ref{doubleslit}) be prepared in the initial state given by a Gaussian with large spread $\sigma$, centred around $0$: 
\begin{equation}
	|\varphi\rangle=\int dy ~\varphi(y)|y\rangle=N\int dy e^{-\frac{y^2}{4\sigma^2}}|y\rangle,
\end{equation} where $N$ is a suitable normalization factor. Together, the particle and the pointer form the compound system with the joint initial state

\begin{equation}
	|\psi\rangle\otimes|\varphi\rangle= \int dxdy ~\psi(x)\varphi(y)|x\rangle\otimes|y\rangle.
\end{equation}

Now consider the first -- the weak -- measurement process. It consists of an interaction between the particle and the pointer. Upon completion of this process (say at $t=0$), the compound system ends up in the entangled state

\begin{equation}
	|\Psi(x,y,0)\rangle= \int dxdy ~\psi(x)\varphi(y-x)|x\rangle\otimes|y\rangle.
\end{equation}

The probability distribution for the pointer variable $y$, \textit{given} some position $X$ of the particle, is therefore:

\begin{equation}
	\rho_X(y)=\frac{|\Psi(X,y,0)|^2}{|\psi(X)|^2}=|\varphi(y-X)|^2.
\end{equation} This probability density determines the expectation value 

\begin{equation}
	\label{expectation value at t=0}
	\mathbb{E}(y|x=X)=\int dy~y \rho_X(y)=X.    
\end{equation} That is, the mean value of the pointer distribution, conditional on the particle occupying position $X$, coincides with that position. This underwrites the following counterfactual:

\begin{itemize}
	\label{C_0}
	\item[$\mathbf{(C_0)}$] \textit{If one were to perform an ordinary (strong) position measurement on the particles \textit{immediately after} the weak interaction, the expectation value of the weak pointer would yield the actual position of the particle.}
\end{itemize} 

Via $\mathbb{E}(y|x=X)$, the particle position is thus empirically accessible through the statistics of large ensembles of identically prepared particles from which we pick the post-selected outcomes $x=X$.  

This thought is further exploited in the final steps of Wiseman's procedure. In the preceding considerations, the strong measurement was performed immediately upon the weak one. Instead, we'll now allow for a small delay. After the particle and the pointer have (weakly) interacted, the total system evolves freely for some small but finite time $\tau$. Its state, then, is    

\begin{equation}
	|\Psi(x,y,\tau)\rangle= e^{-\frac{i}{\hbar}\tau\hat{H}_0}|\Psi(x,y,0)\rangle,
\end{equation} where $\hat{H}_0$ denotes the system’s free Hamiltonian. 

Eventually, we perform a strong measurement of the particle’s position $X_{\tau}$ at $t=\tau$. (The strong coupling between the measurement device and the particle enables precise detection of the latter's actual position.) We thus obtain the expectation value for the pointer variable, conditional on the particle occupying the position $X_{\tau}$ at $t=\tau$:

\begin{equation}
	\label{weak position value}
	\mathbb{E}(y|x=X_{\tau})=\int dy~y |\Psi(X_{\tau},y,\tau)|^2.
\end{equation}

Through the statistics of a sub-ensemble of particles whose strong position measurements yielded $X_{\tau}$, this expectation value is empirically accessible. 

In \textit{analogy} to Equation \eqref{expectation value at t=0}, let's define the position: 

\begin{equation}
	\label{position}
	X_0:=\mathbb{E}(y|x=X_{\tau}).
\end{equation}

Combined with the particle position $X_\tau$, obtained from the strong measurement at $t=\tau$, it thus appears as if we have access to particle positions at two successive moments. Using Equation \eqref{position}, the associated displacement is

\begin{equation}
	\label{displacement}
	X_{\tau}-X_0=X_{\tau}-\mathbb{E}(y|x=X_{\tau}).
\end{equation}

Let's grant one can make it plausible that the particles' trajectories are differentiable. Then, the displacement (Equation \eqref{displacement}) gives rise to the velocity field 

\begin{equation}
	\label{velocity0}
	v(X_0):= \lim\limits_{\tau \rightarrow 0}\frac{1}{\tau} (X_{\tau} - \mathbb{E}(y|x=X_{\tau})).
\end{equation}

Note that all terms on the r.h.s.\ of Equation \eqref{velocity0} are observable. (Hence, presumably, Wiseman's labelling \eqref{velocity0} as an `operational definition'.) In conclusion, it seems, via the statistics of an experimental set-up implementing Wiseman’s procedure, we can empirically probe this velocity field. 

{\em But what does this velocity field signify?} It is tempting to identify it with the particles' actual velocities. That is, should this be true, the flow of Equation \eqref{velocity0} generates the particles' trajectories (assumed to be deterministic and differentiable). Is this identification justified?

By \textit{defining} an $X_0$ via Equation \eqref{position}, our notation certainly suggests so. Let’s assume that this is correct and call this the `Correspondence Assumption' (COR). That is, suppose that the actual particle position $X_{\tau}$ at $t=\tau$ is connected with the earlier particle position $x(0)=X_0=\hat{T}_{-\tau}X_{\tau}$ at $t=0$, where $\hat{T}_{-\tau}$ denotes the shift operator that evolves backwards particle positions by $\tau$. (In other words, for arbitrary initial positions, $\hat{T}_{\tau}$ supplies the full trajectory.) Then, according to (COR), the expectation value (\ref{position}) corresponds to the particles' position at $t=0$. For post-selection of sub-ensembles with $x(\tau)=X_{\tau}$, (COR) thus takes the form (in the limit of large spread $\sigma$)

\begin{equation}
	\textbf{\text{(COR)}}~  \mathbb{E}(y|x(\tau)=X_{\tau})=\hat{T}_{-\tau}X_{\tau}.\footnote{The assumption is in fact what explicitly connects the manifest variables, i.e.\ the accessible records, to the non-manifest variables of the theory, i.e.\ the actual Bohmian particle positions. In other words, the expectation value $\mathbb{E}(y|x=X_{\tau})$ resulting from the weak measurement procedure, is assumed to \textit{signify} particle configurations at a particular time $\tau$.}
\end{equation}

In other words, (COR) implies the counterfactual:

\begin{itemize}
	\label{C_t}
	\item[$\mathbf{(C_t)}$] \textit{If one were to perform a strong position measurement at $t=\tau$ (with the weak interaction taking place at $t=0$), yielding the particles' position at $x(\tau)=X_{\tau}$, the weak value would be directly correlated with the particles' earlier position $\hat{T}_{-\tau}X_{\tau}$. That is, upon a strong measurement at $t=\tau$, the expectation value would reveal the particles' true positions:
		\begin{equation}
			\mathbb{E}(y|x(\tau)=X_{\tau})=\hat{T}_{-\tau}X_{\tau}.
	\end{equation}}
\end{itemize} 

On (COR), the weak value thus gives the particle’s \textit{actual} position at the weak screen: the expectation value on the l.h.s. is reliably correlated with the particle's earlier positions. But most importantly, this is an \textit{if and only if condition}: If (COR) is satisfied, then we recover the actual position; but if it is not, we do not. As a result, one ought to assume that (COR) is valid for weak position measurements to yield actual particle positions. 

Thereby, any set of data compatible with QM appears to corroborate standard de Broglie--Bohm theory: \textit{given} (COR), weak velocity measurements yield standard de Broglie--Bohm theory's velocity field. It thus seems as if standard de Broglie--Bohm theory’s empirical underdetermination has been overcome.

Such an apparent possibility of confirming standard de Broglie--Bohm theory would be remarkable. It crucially hinges, however, on the soundness of (COR). As I shall show (COR) is generically false, and there is no plausible argument for why it should hold. This is eventually illustrated with a simple example.
Prima facie, (COR) looks like a plausible extrapolation of a strong measurement immediately after the weak interaction (i.e.\ at $t=0$). Firstly, (COR) holds in the limit $\tau\rightarrow 0^+$. Next, in a deterministic world, it would seem that

\begin{equation}
	\mathbb{E}(y|x(\tau)=\hat{T}_{\tau}\kappa)=\mathbb{E}(y|x(0)=\kappa),
\end{equation} where $\kappa\in\mathbb{R}$ denotes a position. 

By appeal to $C_0$, this would then yield 

\begin{equation}
	\mathbb{E}(y|x(\tau)=\hat{T}_{\tau}\kappa)=\mathbb{E}(y|x(0)=\kappa)=\kappa,
\end{equation} as desired.

Yet, this inference is illicit. While it is true that $\{(x(\tau),y)\in \mathbb{R}\times\mathbb{R}:x(\tau)=\hat{T}_{\tau}\kappa\}$ and $\{(x(0),y)\in \mathbb{R}\times\mathbb{R}:x(0)=\kappa\}$ contains the same pointer configurations, this \textit{does not} imply that $\mathbb{P}(y|x(\tau)=\hat{T}_{\tau}\kappa)=\mathbb{P}(y|x(0)=\kappa)$. For this to hold, the conditional probabilities -- as defined via post-selection -- on both sides must be well defined. That is, 
\begin{equation}
	\label{conditional probabilities}
	\frac{\mathbb{P}(y \wedge x(\tau)=\hat{T}_{\tau}\kappa)}{\mathbb{P}(x(\tau)=\hat{T}_{\tau}\kappa)} \ \text{and} \ \frac{\mathbb{P}(y \wedge x(0)=\kappa)}{\mathbb{P}(x(0)=\kappa)}
\end{equation} must exist (and coincide). 

In classical statistical mechanics, one may take this for granted. However, in a \textit{quantum} context, entanglement complicates the situation: it compromises the ascription of probability measures to certain events. One must heed the time with respect to which the assigned probability measure is defined. This is the case with weak velocity measurements. Recall that Wiseman’s measurement protocol only performs the strong measurement at $t=\tau$. This precludes defining the second term in \eqref{conditional probabilities}. That is, no strong measurement is performed -- and no attendant `effective collapse' of the wave function occurs -- at an \textit{earlier} time (viz. at $t=0$). As a result, at the time of the weak interaction ($t=0$), the wave function of the pointer and particles is entangled. That means, however, that we \textit{cannot} naively assign the event of any particular particle position at $t=0$ an objective, individual probability measure.

This follows from the fact that $\mathbb{P}(x(0)=\kappa)$ is obtained from the pointer-\textit{cum}-particle system’s reduced density matrix (i.e.\ by partially tracing out the pointer’s degrees of freedom). But this transition from the density matrix of a pure state to the reduced density matrix of an `improper mixture' lacks objective-physical justification (see, e.g.\ Mittelstaedt 2004, Chapters~3 and 4). Contrast that with the situation of $\frac{\mathbb{P}(y \wedge x(\tau)=X_{\tau})}{\mathbb{P}(x(\tau)=X_{\tau})}$: this \textit{is} well defined via post-selection. That is, due to the `effective collapse' (see, e.g.\  D\"urr and Teufel 2009, Chapter 9.2), induced by the strong measurement at $t=\tau$, the event $x(\tau)=X_{\tau}$ \textit{can} be assigned a well-defined probability measure.

No \textit{independent} reasons have been given so far for believing that (COR) is true. Consequently, counterexamples to (COR) abound -- and are perfectly familiar: \textit{any} non-standard variant of de Broglie--Bohm theory of the type of Equation \eqref{altguidanceequation} (i.e.\ with non-vanishing, divergence-free vector field $\textbf{\emph{j}}$). In them, the particle's trajectory generically crosses the weak screen at a point \textit{distinct} from what the weak velocity measurements would make us believe. Fig.\ \ref{fig:correspondence} above illustrates this.

%\begin{figure}[h]
%	\centering
%	\includegraphics[width=.8\textwidth]{double slit.pdf}
%	\caption{The weak measurement procedure for a given post-selected state $x(\tau)=X_{\tau}$. The weak value is obtained from the distribution on the weak screen. When the velocity field is that of standard de Broglie--Bohm theory ($\textbf{\emph{j}}=0$), the actual position of the particle $x(0)$ matches the weak value $x_w$. For an alternative guidance equation ($\textbf{\emph{j}}\neq 0$), it does not: the particle crosses the weak screen at a point $x'(0)$, other than the weak value. This shows that depending on which guidance equation one chooses, the weak value need not yield the actual position of the particle at time $0$.}
%	\label{fig:correspondence}
%\end{figure}

The outcome cannot be considered as representing the actual position of the particle at time $t$. It is just unknown: it could have traversed \textit{any} location within the support of the Gaussian wave function, centered around the weak value. Still, the operationally defined velocity (obtained from averaging) would not change: as long as the Born rule and the validity of the Schrödinger equation hold, its value remains the same. (In this sense, any guidance equation of the type of Equation \eqref{altguidanceequation}, is compatible with Wiseman's operationally defined velocity.)  

In the absence of an independent argument for the correspondence between weakly measured and actual positions (i.e.\ COR), it remains unclear what -- if anything -- Wiseman's operational velocity \eqref{operational velocity} signifies non-manifestly, i.e.\ ontologically.

The assumption $C_t$, necessary for the correspondence of weak and actual velocities, is, in fact, equivalent -- in virtue solely of the quantum mechanical formalism and the supposition of deterministic differentiable particle trajectories -- to standard de Broglie--Bohm theory. (Firstly, suppose that $C_t$ is true. Then, the weak velocity measurement yields the actual particle velocities. Wiseman's operationally defined velocity \eqref{operational velocity} uniquely picks out a guidance equation---that of standard de Broglie--Bohm theory.\footnote{On a related note, D\"urr, Goldstein, and Zangh\'i identify a characteristic feature of standard pilot-wave theory's velocity field as the `crucial condition' for why, allegedly, Wiseman's measurement protocol constitutes `genuine' measurements of Bohmian trajectories (D\"urr {\em et al}. 2013, Chapter 7; see also D\"urr {\em et al}.\ 2013, Chapter 3, and Goldstein 2017, Section 4 for similar claims). The assumption effectively amounts to the fact that in the standard guidance equation a particle's velocity field only depends on the particle's wave function whenever the particle-\textit{cum}-pointer compound system has the form $\psi(x)\otimes\phi(y-x)$. It uniquely singles out standard de Broglie--Bohm theory. However, DGZ seem to misidentify that `crucial condition'. Indeed, it plays no obvious role in the attempt to exploit weak velocity measurements for the standard guidance equation, and nowhere it is invoked explicitly (Fankhauser and D\"urr 2021).}

Conversely, suppose standard de Broglie--Bohm theory to be true. A weak velocity measurement then discloses the actual particle velocities. Thus, $C_t$ holds.) 

\subsection{Non-empirical support for pilot-wave theory?}
\label{grounding}

The main result of Wiseman's paper can be considered as a conditional claim: \textit{if} one adopts his operationally defined velocity (and assumes deterministic, differential particle trajectories), it is uniquely determined as that of standard de Broglie--Bohm theory; on this reading, Wiseman remains neutral -- whether it is plausibly satisfied (or not).   

Wiseman's envisioned reasoning pertains to a non-empirical motivation for standard pilot-wave theory: Albeit not per se referring to individual particles, the statistically defined operational velocity provides a `justification for [standard de Broglie--Bohm theory's] law of motion [i.e.\ the standard guidance equation]' (Wiseman 2007, p.~2). This in turn purports that the theory's standard form `is preferred over all other on physical grounds' (Wiseman 2007, p.~12). That is, although other velocity fields generate the same (statistically empirically accessible) mean velocity, we ought to believe that the standard velocity field is true -- rather than any of its alternatives.

Similarly, D\"urr \emph{et al}. turn on the allegedly natural character of his proposal to operationally define velocities via weak values: `(Standard de Broglie--Bohm theory) delivers thus the most natural explanation of the experiments described' (D\"urr and Lazarovici 2018, p.~145, my translation). 

But as I have illustrated in Section 9.3.2, weak positions do not reveal the actual particle position. What, then, is the proposed relationship between weak velocities and actual particle velocities supposed to be?

The disadvantages of naive realism about weak position values are demonstrated in the so-called Three-Box Paradox (Aharonov and Vaidman 1991; Aharanov and Rohrlich 2005, Chapter 16.5; Maroney 2017). Here, we encounter another illustrative case where the identification of the manifest with the non-manifest is more complex than one would hope.

Imagine a particle and three boxes labelled $A$,$B$, and $C$. Let the particle's initial state be 

\begin{equation}
	|\psi_{in}\rangle= \frac{1}{\sqrt{3}}(|A\rangle+|B\rangle+|C\rangle),
\end{equation} where $|A\rangle$ denotes the state in which the particle is in box $A$, and similarly, $|B\rangle$ and $|C\rangle$. For its final state, on which we post-select, choose

\begin{equation}
	|\psi_{fin}\rangle= \frac{1}{\sqrt{3}}(|A\rangle+|B\rangle-|C\rangle).
\end{equation}

Via the definition of weak values, one then obtains the resulting weak values for the projectors onto state $i\in {A,B,C}$, $\hat{P}_i:=|i\rangle\langle i|$:

\begin{align}
	\langle\hat{P}_A\rangle_w&= 1\nonumber\\
	\langle\hat{P}_B\rangle_w&= 1\nonumber\\
	\langle\hat{P}_C\rangle_w&= -1.
\end{align} 

If one were to believe that weak values invariably reveal the real positions of particles, one would have to conclude that box $C$ contains $-1$ particles. However, in de Broglie--Bohm theory's ontology (in any of its variants), particles either occupy a position or do not occupy a position.

Consequently, Bohmians must be careful of interpreting weak values as real position values without qualification.

We are thus left with, at best, a considerably weaker position, one close to Bricmont (2016, p.~136): `[Weak velocity measurements via Wiseman’s protocol] (are) not meant to `prove' that the de-Broglie-Bohm theory is correct', because other theories will make the same predictions. Still, the result is nevertheless suggestive because the predictions made here by the de Broglie--Bohm theory are very natural within that theory [...].'

\section[Surrealistic trajectories]{Declaring Bohmian trajectories as surreal: falsifying Bohmian predictions?}
\sectionmark{Surrealistic Trajectories}
\label{section:Surrealistic Bohmian trajectories}

The detection procedure leading to what has (somewhat inappropriately) been coined `\textit{surrealistic}' trajectories presents another particularly illustrative example where a naive conflation of the manifest and non-manifest leads to inconsistencies. (The term `surrealistic' is misleading as there is nothing in the standard quantum theory that one would refer to as being \textit{real}.) Experimental proposals to measure such trajectories were deemed to rule out pilot-wave theory and present a second attempt to detect, or rather falsify, Bohmian trajectories. We observe that the standard QM lacks a foundation for when a trajectory is supposed to be \textit{measurable}. This, in fact, would block the paradox right from the outset because any claim that some `path of a particle' was detected in an experiment is unwarranted. 

There is no need to repeat the many treatments of this experiment in the literature. Instead, the point here is to present surrealistic trajectories as a case where a consistent underpinning of quantum theory leads to the fact that not all position measurements are genuine measurements of a particle's actual position. Thus, such experiments \textit{cannot} falsify the existence of pilot-wave trajectories.  

\subsection{The surrealistic trajectories experiment}
\label{section:surrealistic trajectories experiment}

\begin{figure}[h]
	\centering
	\includegraphics[width=0.3\linewidth]{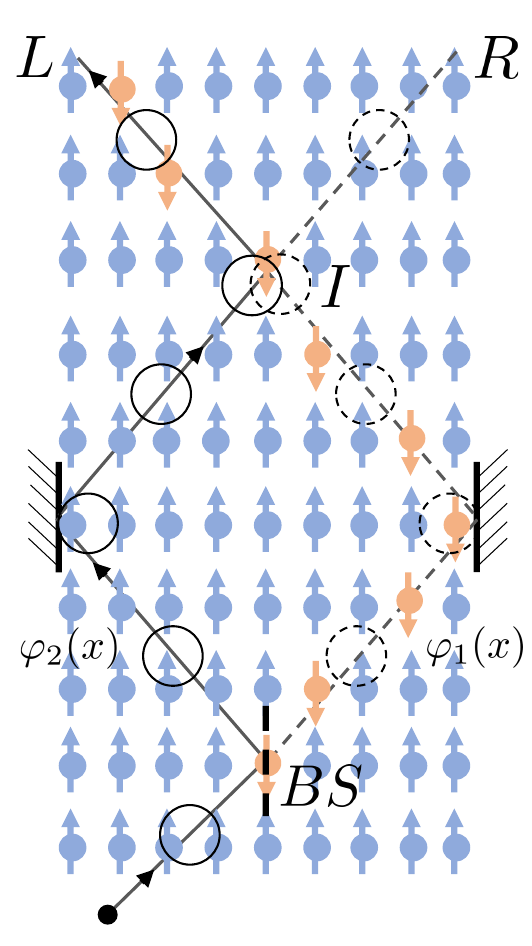}
	\caption{A particle's trajectory is `detected' by a grid of spins that flip when the particle travels by them. Although both theories produce the same outcomes for the spin flips, as opposed to standard QM, pilot-wave theory predicts the particle to traverse a path that is not indicated by the series of spin flips (the solid black $L$ path to the left).}
	\label{fig:spin-flip-surreal}
\end{figure}

In the surreal trajectory experiment -- an instructive variant of which I discuss here\footnote{Another simple account is discussed in \cite{barrett2000persistence}, and the original experiment was presented in \cite{EnglertScullySussmannWalther}.} -- a Mach--Zehnder interferometer is supplemented with a grid of spin systems interacting with the particle. The interaction is designed such that when the particle traversing the interferometer hits, the corresponding spins flip from their ready state $|\uparrow\rangle$ to the state $|\downarrow\rangle$. This allows, supposedly, to track the particle's path. After the experiment is performed, the spin flips are read off, indicating the trajectory the particle took through the experiment (see the set-up in Fig.\ \ref{fig:spin-flip-surreal}).  

Assume that before the interaction the experiment consists of $n$ unperturbed spins $|\uparrow\rangle^{\otimes n}$. Without loss of generality, we assign the spins in the path $L$ to the first $k$ slots of the total state and the spins in the alternative path $R$ to the next $k$ slots, respectively. Hence, the state is represented as

\begin{equation}
	|\uparrow\rangle^{\otimes n}:=|\uparrow_{L_1}\rangle \cdots |\uparrow_{L_k}\rangle|\uparrow_{R_1}\rangle \cdots |\uparrow_{R_k}\rangle |\uparrow\rangle^{\otimes (n-2k)}.
\end{equation}

The quantum description of the experiment looks like so: when the particle travels through the experiment, the total initial state of particle and spin grid  
\begin{equation}
	\label{eqn:initial state surreal trajectory}
	\frac{1}{\sqrt{2}}(|\varphi_1(x)\rangle+|\varphi_2(x)\rangle)|\uparrow\rangle^{\otimes n}
\end{equation} evolves into the final post-interaction state

\begin{equation}
	\label{eqn:post measurement spin flips surreal trajectory}
	\frac{1}{\sqrt{2}}(|\varphi_1'(x)\rangle |\downarrow_{L_1}\rangle \cdots |\downarrow_{L_k}\rangle |\uparrow\rangle^{\otimes (n-k)}+ |\varphi_2'(x)\rangle|\uparrow\rangle^{\otimes k}|\downarrow_{R_1}\rangle \cdots |\downarrow_{R_k}\rangle |\uparrow\rangle^{\otimes (n-2k)}),
\end{equation} where $|\varphi_1(x)\rangle, |\varphi_2(x)\rangle$ are the two disjoint wave packets of the particle emerging from the beam-splitter BS, and $|\varphi_1'(x)\rangle, |\varphi_2'(x)\rangle$ are the two disjoint wave packets of the particle at the end of the experiment (top region in Fig.\ \ref{fig:spin-flip-surreal}). 

The two branches of the wave function are reflected at two mirrors and subsequently overlap in the region $I$ before they travel further on their path. In light of the standard quantum description, it is clear what is going on in terms of how the trace of spin flips arises. Each branch of the particle's wave function causes a series of adjacent spin particles to switch states like a game of dominoes, creating a path a single particle would be expected to take. Since the initial quantum state (Equation \eqref{eqn:initial state surreal trajectory}) is a superposition of those two branches, each track occurs with equal probability. Admittedly, this looks daringly close to the detection of a particle's trajectory. Nothing mysterious so far. 

But pilot-wave theory (committed to actual particle positions) tells a different story of where the particle went. To compute the Bohmian prediction, the total wave function of all involved particles is rewritten in position space, and spin degrees of freedom are used in spinor form. The trajectory is then calculated from Bohm's guiding equation.

The total quantum state of an individual spin particle on the grid then reads

\begin{equation}
	\Phi(y)\begin{pmatrix}
		1 \\
		0		
	\end{pmatrix},
\end{equation} for spin up, and 
\begin{equation}
	\Phi(y) \begin{pmatrix}
		0 \\
		1		
	\end{pmatrix},
\end{equation} for spin down, respectively. Here, $\Phi(y)$ is a single spin particle's spatial wave function centred at the position of the particle. Note, importantly, that the wave function of the position degree of freedom is identical for both spin values. 

The total final state, thus, is found to be 

\begin{equation}
	\begin{split}
		\frac{1}{\sqrt{2}} \bigg(|\varphi_1'(x)\rangle\Phi_{L_1}(y-y_{L_1})\begin{pmatrix}
			0 \\
			1		
		\end{pmatrix} \cdots \Phi_{L_k}(y-y_{L_k})\begin{pmatrix}
			0 \\
			1		
		\end{pmatrix}
		\\	
		\otimes \Phi_{R_1}(y-y_{R_1})\begin{pmatrix}
			1 \\
			0		
		\end{pmatrix} \cdots \Phi_{R_k}(y-y_{R_k})\begin{pmatrix}
			1 \\
			0		
		\end{pmatrix} |\uparrow\rangle^{\otimes(n-2k)} 
		\\
		+|\varphi_2'(x)\rangle \Phi_{L_1}(y-y_{L_1})\begin{pmatrix}
			1 \\
			0		
		\end{pmatrix} \cdots \Phi_{L_k}(y-y_{L_k})\begin{pmatrix}
			1 \\
			0		
		\end{pmatrix}
		\\
		\otimes \Phi_{R_1}(y-y_{R_1})\begin{pmatrix}
			0\\
			1		
		\end{pmatrix} \cdots \Phi_{R_k}(y-y_{R_k})\begin{pmatrix}
			0 \\
			1		
		\end{pmatrix} |\uparrow\rangle^{\otimes(n-2k)}\bigg)
	\end{split}
\end{equation} with $y_{L_i}, y_{R_i}$ being the locations of the spin particles. Nothing mysterious so far here either. But what happens in the region $I$ according to the Bohmian description? 

Throughout the experiment, the spatial degrees of freedom of the spin particles remain untouched since only their spin state flips. In other words, the spin measurement is in this sense incomplete since no record was created in terms of manifest configurations that would displace the spin particles (or further entangle them to spatial degrees of freedom of an apparatus). 

Therefore, without a manifest read-out, each particle's spatial wave function is the same irrespective of whether that particle's spin has flipped or not, such that they can be pulled out from the superposition of the final state, i.e.

\begin{equation}
	\frac{1}{\sqrt{2}}\prod_{i=1}^{k}\Phi_{L_i}\Phi_{R_i}\left(|\varphi_1'(x)\rangle\begin{pmatrix}
		0 \\
		1		
	\end{pmatrix}^{\otimes k} |\uparrow\rangle^{\otimes (n-k)}+|\varphi_2'(x)\rangle |\uparrow\rangle^{\otimes k}\begin{pmatrix}
		0 \\
		1		
	\end{pmatrix}^{\otimes k}|\uparrow\rangle^{\otimes (n-2k)}\right).
\end{equation} 

This makes it possible for the two branches of the target particle's wave function to interfere coherently in region $I$. Plugging the state in $I$ into the Bohmian guiding equation for spinor-valued wave functions reveals that the target particle's velocity has contributions from both branches of its wave function, i.e.\

\begin{equation}
	\label{eqn:surreal wave function}
	m v_x= \hbar\Im \frac{\Psi^*\nabla_x \Psi}{\Psi^*\Psi}	\sim ((\varphi^I_1(x))^*\nabla\varphi_1^I(x)+(\varphi^I_2(x))^*\nabla\varphi_2^I(x)),
\end{equation}
\noindent where $\Psi$ is the total wave function of the system.
The guiding potential of particle $x$ is independent of all $y$ degrees of freedom. The interaction is designed such that only the spin state evolves when a particle passes by, and the spin particles do not move upon that (otherwise the effect would not occur). If one of the spin particles were to move during the interaction, one of the two contributions would be suppressed. That is, although the two spin states (the left and right branches) of an individual spin particle are orthogonal, i.e.\ $\langle\Phi(y)|\Phi(y)\rangle \langle\uparrow|\downarrow\rangle=0$, they may still overlap in configuration space. Hence, the overlapping spin states in configuration space will affect the target particle's motion as Equation \eqref{eqn:surreal wave function} suggests.\footnote{This was highlighted in \cite{maroney-phd-thesis-2004} by distinguishing orthogonality of quantum states from `super-orthogonality', where the latter refers to the feature that two quantum wave functions do not overlap in configurations space.}  

Equation \eqref{eqn:surreal wave function} shows that the particle will move straight up when the two branches are made to overlap since the total velocity is the sum of the two contributions coming from the left and right moving wave packet (i.e.\ horizontal velocity component cancels). When the particle reaches the interference area $I$, the two wave packets can interfere before they continue to pass through each other and separate again. Moreover, this causes the particle to be reflected from point $I$ and continue travelling on either the experiment's left or right half.   

In the last step, a measurement of the final position of the particle is performed, and the spin states are recorded (e.g.\ with a series of Stern--Gerlach apparatuses). Note that if the spatial wave functions and positions $y_i$ of the spin particles on the grid were to change, they could not be pulled out as an overall factor in Equation \eqref{eqn:surreal wave function}, and would therefore effectively decohere the target particle's wave function into its two branches.

\subsection{What surrealistic trajectories signify}
\label{section:significance of surrealistic trajectories}

As a result, the standard quantum framework predicts a chain of spin flips to emerge that \textit{does not} coincide with the particle trajectory as predicted by the Bohmian guiding equation. If we believe that the spin track in the experiment reliably indicates the particle's path, one could elicit a puzzle: the observed path (the flipped spins) is the one on the right, but according to de Broglie--Bohm theory, the particle was never there. That is, the Bohmian trajectory is `surreal', and therefore the theory cannot be right.

We first observe that, of course, concerning the state of the spin, whether flipped or unflipped, the hypothetical emergence of a chain of flipped spins is not an actual measurement of anything yet. Although one may claim that some spin state must have changed upon interaction with the particle flying through the experiments, no configurational record exists in the manifest domain indicating a measurement result. This turned out to be crucial. The argument relies on the fact that none of the spin particle's configurations $y_i$ change during the experiment. Otherwise, the guiding equation would not give rise to the described behaviour of the target particle. But this also means that during the experiment, the spin chain must not be observed in terms of manifest configurations. After the interference region $I$ was passed, subsequent measurements may create records thereof. 

The existence of the so-called surrealistic trajectories was supposed to be evidence for testable false predictions of pilot-wave theory (see Englert {\em et al}. 1992). The trajectories predicted, so the story goes, contradict what an actual measurement of the particle trajectory would yield. So, they were used to claim that the Bohmian theory gives wrong predictions for the \textit{actual} position of a particle. If this were true, it would falsify the theory.

However, one firstly recalls that, the predictions of de Broglie--Bohm theory for the measurement outcomes of this thought experiment coincide with the ones of standard QM. Secondly, in the standard theory, no meaning is assigned to the trajectory of a particle so that it remains unclear what such a measurement would even have to do with the dynamics of a particle (cf.\ D\"urr {\em et al}. 1993; Lazarovici 2020; Bricmont 2016; D\"urr {\em et al}. 2013, Chapter~3). Classical particles do not exist in standard quantum theory, let alone have trajectories. As I alluded, `surrealistic' trajectories result from an unjustified relationship between the manifest and non-manifest domains. The observed spin track is naively identified with a particle's actual trajectory. Since this claim would go beyond the postulates of standard QM, it is unclear whether that assumption can be consistently made. Furthermore, a theory where a claim about the existence of actual particle trajectories \textit{can} consistently be made is pilot-wave theory. But as the example shows, the trajectory cannot coincide with the one indicated by the chain of spin flips. I therefore conclude that if particle trajectories are assumed to exist in quantum theory, not all `position' measurements can be interpreted as measurements of a particle's actual position.

There are other analogous instances of `surreal' quantum observables. For example, as Brown \emph{et al.} (1995) indicate, other properties besides spin, such as mass, charge, and magnetic moment are all inconsistent with what they call the thesis of `localized particle properties'. 

\section{Conclusions}
\label{conclusion}

In the first part of this chapter, weak velocity measurements of Bohmian particles are analyzed. I have explained the status of weak velocity measurements in two ways. On the one hand, they are an interesting application of standard QM in a novel experimental regime (viz., that of weak pointer-system couplings). They allow us to empirically probe the gradient of the system's wave function -- irrespective of any particular interpretation of the quantum formalism. On the other hand, however, with respect to the significance of weak velocity measurements, weak velocity measurements shed no light on the status of standard de Broglie--Bohm theory. In particular, on their own, weak velocity measurements do not provide any convincing support -- empirical or non-empirical -- for standard de Broglie--Bohm theory over any of its alternative versions. The upshot of the analysis by which the puzzle resolves is that one should not naively identify weak positions with actual Bohmian positions. As a result, this highlights the importance of defining the manifest variables of a theory, i.e.\ the experimentally accessible records, from the non-manifest variables, i.e.\ ontological variables -- the Bohmian positions. The Three-Box Paradox demonstrated the dangers of any naive realism about weak position values. 

The discussion continued with the allegedly paradoxical claims commonly made on the so-called `surrealistic' trajectories in pilot-wave theory, where detectors are purportedly misled. This example served as an another instructive case for inconsistent reasoning on what quantum outcomes are supposed to signify. The analysis showed that a detection of `surrealistic' trajectories provides no empirical argument against the Bohmian predictions for particle trajectories. Therefore, both in the same vein, neither surrealistic trajectories nor weak velocity measurements qualify as falsification or confirmation in favour of the velocity field, postulated by standard de Broglie--Bohm theory in any robust sense. 

While certain advocates of pilot-wave theory were somewhat assertive in claiming that weak measurements substantiated the theory or specific variants of it, similarly, some opponents of the theory have been too assertive. They contend that alternative indirect methods aimed at empirically inferring information about particle trajectories contradict the theory. The fundamental issue lies in the absence of a particle trajectory concept within the standard quantum framework, and the central insight is that quantum measurements have no straightforward bearing on variables in the non-manifest domain.

\end{document}